\documentclass[a4paper,fleqn]{cas-sc}

\usepackage[numbers]{natbib}

\def\tsc#1{\csdef{#1}{\textsc{\lowercase{#1}}\xspace}}
\tsc{WGM}
\tsc{QE}
\tsc{EP}
\tsc{PMS}
\tsc{BEC}
\tsc{DE}

\begin{document}
\let\WriteBookmarks\relax
\def\floatpagepagefraction{1}
\def\textpagefraction{.001}
\shorttitle{ATHIC proceedings}
\shortauthors{Xin-Li Sheng et~al.}

\title [mode = title]{Spin alignment of vector mesons in heavy-ion collision}                                          
\tnotemark[1]

\tnotetext[1]{X.L.S. is supported by the Italian Ministry of University and Research, project PRIN2022 ”Advanced probes of the Quark Gluon Plasma”, funded by Next Generation EU, Mission 4 Component 1.}

\author[1]{Xin-Li Sheng}[style=chinese]
\cormark[1] 
\ead{sheng@fi.infn.it}
\affiliation[1]{organization={Università degli studi di Firenze and INFN Sezione di Firenze},
            addressline={Via G. Sansone 1}, 
            city={Sesto Fiorentino (Florence)},
            postcode={I-50019}, 
            state={},
            country={Italy}}
            
\author[2,3]{Qun Wang}[style=chinese]
\ead{qunwang@ustc.edu.cn}
\affiliation[2]{organization={Department of Modern Physics, University of Science and Technology of China},
            city={Hefei, Anhui},
            postcode={230026}, 
            country={China}}
\affiliation[3]{organization={School of Mechanics and Physics, Anhui University of Science and Technology},
            city={Huainan, Anhui},
            postcode={232001}, 
            country={China}}

\cortext[cor1]{Corresponding author}

\begin{abstract}
We give a brief review on the spin alignment induced by the strong field and the shear-stress tensor. In experiments, a significant positive deviation from 1/3 is observed for the $\phi$ meson, which can be explained by the anisotropy of the strong field fluctuation in the meson's rest frame, while the anisotropy is mainly a consequence of the motion of meson relative to the quark-gluon plasma. On the other hand, the shear-induced spin alignment is of the order $10^{-4}\sim10^{-5}$ if the magnitude of thermal shear tensor is $10^{-2}$.
\end{abstract}


\maketitle
\section{Introduction}

Relativistic heavy-ion collisions generate strongly interacting matter called the quark-gluon plasma (QGP). Due to the large initial orbital angular momentum carried by the two colliding nuclei, the system generates an extremely strong vorticity field, which can polarize quarks and then lead to spin polarizations of hyperons observed in experiments through weak decays \cite{Liang:2004ph,STAR:2017ckg}. On the other hand, the polarized quarks can also form vector mesons with nontrivial spin alignment, which can be measured through the polar angle distribution in strong $p$-wave decays. Such an effect was firstly proposed by Liang and Wang \cite{Liang:2004xn} and was confirmed in Au+Au collisions by the STAR collaboration in 2022 \cite{STAR:2022fan}. For the $\phi$ mesons, the global spin alignment shows a significant positive deviation from 1/3, which increases at lower energies. 

From the theory side, the large spin alignment for the $\phi$ meson cannot be accounted by conventional mechanisms like the vorticity or magnetic fields \cite{Yang:2017sdk, Sheng:2020ghv}. Recently, we proposed that the strong force field could be the dominant reason for the spin alignment. In the quark-gluon plasma, this field has nearly vanishing mean value but exhibits a large fluctuation. Using a relativistic quark coalescence model, we derive the relation between the spin alignment for the vector meson and the spin-spin correlation between the constituent quark and antiquark, which corresponds to the fluctuation of strong force field within the scale of meson size \cite{Sheng:2022wsy, Sheng:2022ffb}. The main results and conclusions will be reviewed in Sec. \ref{sec:SFF}. On the other hand, the spin alignment could also contain a linear response to the thermal shear tensor \cite{Li:2022vmb,Wagner:2022gza,Dong:2023cng}, which is first order in a scheme of hydrodynamic gradient expansion and thus could has a sizable contribution. The shear-induced spin alignment will be reviewed in Sec. \ref{sec:shear}. One can also read Refs. \cite{Becattini:2022zvf,Chen:2023hnb,Becattini:2024uha,Chen:2024afy} for recent reviews on spin polarization and spin alignment in heavy-ion collisions.

\section{Spin alignment induced by strong force field}\label{sec:SFF}

In a quark coalescence model, the spin alignment of a vector meson is fully determined by the spin polarizations of its constituent quark and antiquark, while the relative orbital angular momentum between quark and antiqaurk is neglected. We focus on the $\phi$ meson, which consists of $s$ and $\bar{s}$. Assuming that $s$ and $\bar{s}$ are at thermal equilibrium, their spin polarizations are given by \cite{Sheng:2022wsy,Sheng:2022ffb},
\begin{equation}
    P_{s/\bar{s}}^\mu (x, \mathbf{p}) \approx \frac{1}{4m_s} \epsilon^{\mu \nu \rho \sigma} p_\nu \left[ \omega_{\rho \sigma} \pm \frac{Q_s}{(u \cdot p) T} F_{\rho \sigma} \pm \frac{g_\phi}{(u \cdot p) T} F_{\rho \sigma}^\phi \right]\,,
\end{equation}
where $T$ denotes the temperature at the hadronization stage, $m_s$ is the $s$-quark's mass, $u^\mu$ is the local four-velocity of the thermal medium. 
The terms on the right-hand-side denote contributions from the vorticity field $\omega^{\mu\nu}$, the classical electromagnetic field $F^{\mu\nu}$, and the long-wavelength components of the strong force field (the vector $\phi$ field $F_\phi^{\mu\nu}$ in this case), respectively. The vector $\phi$ field captures the strong interaction between strange quarks in the surrounding medium and the constituent quarks inside the $\phi$ meson. The contribution of vector $\phi$ field has a similar form as the classical electromagnetic field, while the corresponding coupling constant $\alpha_\phi=g_\phi^2/(4\pi)\sim \mathcal{O}(1)$ is much larger than $\alpha_\text{EM}=e^2/(4\pi)\approx1/137$. A detailed calculation \cite{Sheng:2019kmk} shows that the contribution from electromagnetic field to the spin alignment is $\mathcal{O}(10^{-3}\sim 10^{-4})$, much smaller than what was observed in experiments \cite{STAR:2022fan}. The vector $\phi$ field, due to a larger coupling constant, is expected to have a much larger contribution than the electromagnetic field. In later discussions, we will neglect the influence of electromagnetic field, which can be easily reproduced from the effects of vector $\phi$ field by taking replacements $g_s\rightarrow Q_s$ and $F^\phi_{\mu\nu}\rightarrow F_{\mu\nu}$.

In Ref. \cite{Sheng:2022ffb}, we present a relativistic kinetic theory for vector mesons, which is derived from the Kadanoff-Baym equation in the closed-time-path formalism. By solving the kinetic theory in a dilute gas limit, we obtain a solution for the spin density matrix of vector mesons created through quark-antiquark coalescence. For the $\phi$ meson, the density matrix is given by
\begin{align}
\rho_{\lambda_1\lambda_2}^\phi(x,{\bf p})=&\frac{\Delta t}{32}\int\frac{d^3{\bf p}^\prime}{(2\pi)^2}\frac{1}{E_{\bf p}^\phi E^{\bar{s}}_{{\bf p}^\prime} E^s_{{\bf p}-{\bf p}^\prime}}f_{\bar{s}}(x,{\bf p}^\prime)f_s(x,{\bf p}-{\bf p}^\prime) \delta\left(E_{\bf p}^\phi-E^{\bar{s}}_{{\bf p}^\prime}- E^s_{{\bf p}-{\bf p}^\prime}\right)\epsilon_\alpha^\ast(\lambda_1,{\bf p})\epsilon_\beta(\lambda_2,{\bf p}) \nonumber\\
&\times \text{Tr}\left\{\gamma^\beta(p^\prime\cdot\gamma-m_s)\left[1+\gamma_5\gamma\cdot P_{\bar{s}}(x,{\bf p}^\prime)\right]\gamma^\alpha\left[(p-p^\prime)\cdot\gamma+m_s\right]\left[1+\gamma_5\gamma\cdot P_s(x,{\bf p}-{\bf p}^\prime)\right]\right\}\,, \label{eq:density_matrix}
\end{align}
where $E_{\bf p}^\phi,\,E^{\bar{s}}_{{\bf p}^\prime},\,E^s_{{\bf p}-{\bf p}^\prime}$ denote the on-shell energies for $\phi$ meson, $s$ quark, and $\bar{s}$ antiquark, respectively, $\Delta t$ denotes the formation time of the meson, $f_{s/\bar{s}}$ are Fermi-Dirac distributions at thermal equilibrium, and $\epsilon^\mu(\lambda,{\bf p})$ is the polarization vector for the $\phi$ meson. We note that the density matrix given in Eq. (\ref{eq:density_matrix}) has not yet been properly normalized. By redefining $\bar{\rho}^\phi_{\lambda_1\lambda_2}\equiv\rho^\phi_{\lambda_1\lambda_2}/\text{tr}\rho^\phi$, we obtain the normalized density matrix in the phase space, where ``$\text{tr}$'' represents the trace in the spin space. Substituting the quark/antiquark polarizations into Eq. (\ref{eq:density_matrix}) and completing the momentum integral in the rest frame of the $\phi$ meson, we derive a simple expression for the spin alignment of the $\phi$ meson, 
\begin{align}
\bar\rho^\phi_{00} (x,p) \approx 
&\frac{1}{3}  + C_1\left[\frac{1}{3} \boldsymbol{\omega}^\prime\cdot\boldsymbol{\omega}^\prime-(\boldsymbol{\epsilon}_{0} \cdot \boldsymbol{\omega}^\prime)^2\right]+ C_2\left[\frac{1}{3} \boldsymbol{\varepsilon}^\prime\cdot\boldsymbol{\varepsilon}^\prime-(\boldsymbol{\epsilon}_{0} \cdot \boldsymbol{\varepsilon}^\prime)^2\right]\nonumber \\
&- \frac{4g_{\phi}^{2}}{m_{\phi}^{2}T^{2}} C_{1} 
\left[ \frac{1}{3} {\bf B}^\prime_{\phi} \cdot {\bf B}^\prime_{\phi} - (\boldsymbol{\epsilon}_{0} \cdot {\bf B}^\prime_{\phi})^{2} \right] 
 - \frac{4g_{\phi}^{2}}{m_{\phi}^{2}T^{2}} C_{2} 
\left[ \frac{1}{3} {\bf E}^\prime_{\phi} \cdot {\bf E}^\prime_{\phi} - (\boldsymbol\epsilon_{0} \cdot {\bf E}^\prime_{\phi})^{2} \right], \label{spin-alignment}
\end{align}
where $T$ is the temperature at the hadronization stage, $\boldsymbol{\omega}^\prime$ and $\boldsymbol{\varepsilon}^\prime$ are thermal vorticity and thermal acceleration, ${\bf E}_{\phi}^\prime$ and ${\bf B}_{\phi}^\prime$ denote the electric and magnetic parts of the vector $\phi$ field tensor. The spin quantization direction is denoted by $\boldsymbol{\epsilon}_0$, which is set to $(0,1,0)$ when measuring the global spin alignment. In Eq. (\ref{spin-alignment}), all fields with primes are defined in the meson's rest frame, which are related to those in the lab frame as, for example,
\begin{align}
{\bf B}^\prime_\phi=&\gamma{\bf B}_\phi-\gamma{\bf v}\times{\bf E}_\phi+(1+\gamma)\frac{{\bf v}\cdot{\bf B}_\phi}{v^2}{\bf v}\,,\nonumber\\
{\bf E}^\prime_\phi=&\gamma{\bf E}_\phi+\gamma{\bf v}\times{\bf B}_\phi+(1+\gamma)\frac{{\bf v}\cdot{\bf E}_\phi}{v^2}{\bf v}\,,
\end{align}
where $\gamma=E_{\bf p}^\phi/m_\phi$ is the Lorentz factor and ${\bf v}={\bf p}/E_{\bf p}^\phi$ is the velocity of the $\phi$ meson. The coefficients $C_{1}$ and $C_{2}$ in Eq. (\ref{spin-alignment}) only depend on masses of the $\phi$ meson and the $s$ quark, whose explicit expressions are given in Ref. \cite{Sheng:2022wsy,Sheng:2022ffb}. From Eq. (\ref{spin-alignment}), we know that the spin alignment depends on the anisotropy of field fluctuations in the meson’s rest frame. In heavy-ion collisions, such an anisotropy may arise from a faster longitudinal expansion of the quark-gluon plasma than the transverse expansion, or from the motion of mesons relative to the background. In the later case, the Lorentz boost amplifies the transverse field components while keeps the longitudinal components invariant, leading to an anisotropy along the direction of motion.

\begin{figure}[h]
    \centering
    \includegraphics[width=0.4\linewidth]{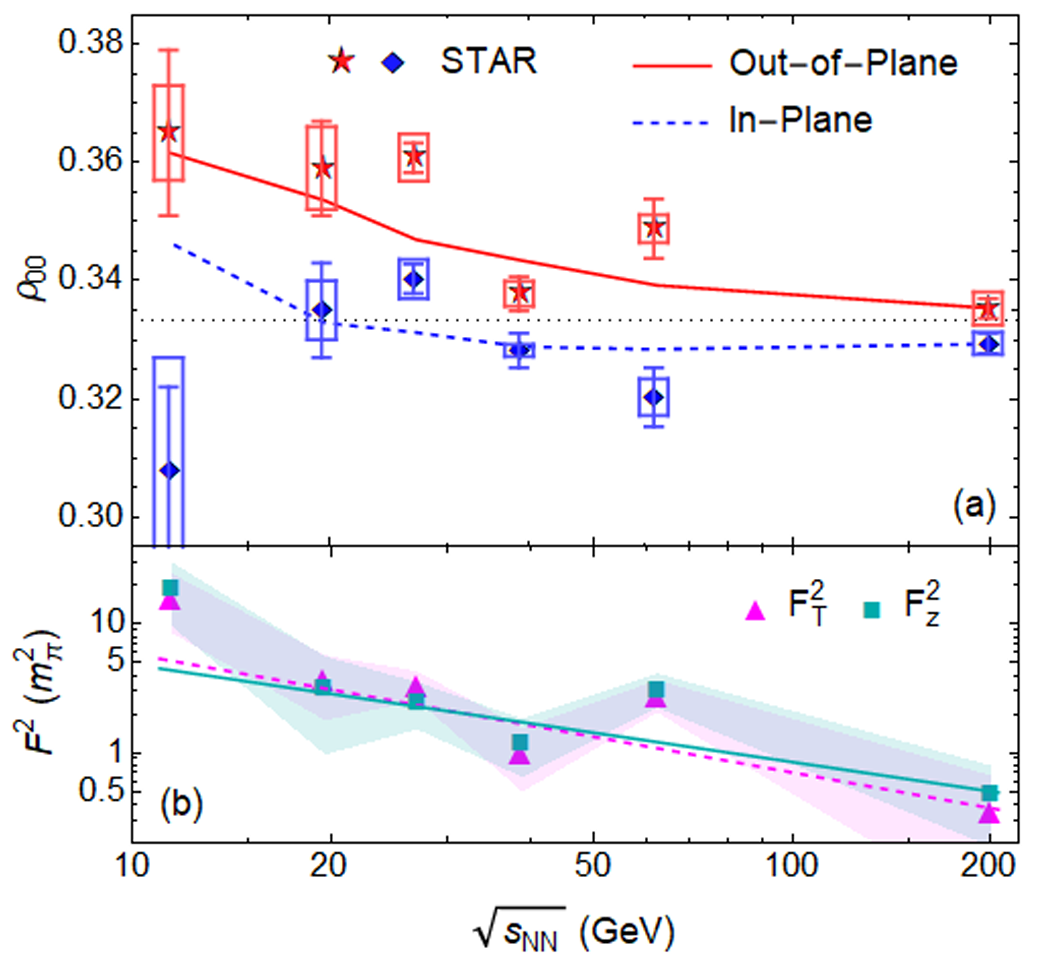}
    \caption{(a): The STAR's experimental data \cite{STAR:2022fan} and our theory prediction \cite{Sheng:2022wsy} on the $\phi$ meson's spin alignment in the out-of-plane and the in-plane directions in Au+Au collisions, as functions of collision energies. (b): Parameters $F_{T}^{2}$ (magenta triangles) and $F_{T}^{2}$ (cyan squares) extracted from the STAR's data. See Ref. \cite{Sheng:2022wsy} for details.}
    \label{fig:rho-00-eng}
\end{figure}

Due to the lack of theoretical inputs on the magnitudes of fluctuations, we treat them as parameters. Numerical simulations show that contributions from vorticity and acceleration are very small and thus we focus on the vector $\phi$ field. Considering the geometry of the quark-gluon plasma, we choose two independent parameters, 
$\langle(g_{\phi}B_{x,y}^{\phi}/T_{h})^{2}\rangle = \langle(g_{\phi}E_{x,y}^{\phi}/T_{h})^{2}\rangle \equiv F_{T}^{2}$ and
$\langle(g_{\phi}B_{z}^{\phi}/T_{h})^{2}\rangle = \langle(g_{\phi}E_{z}^{\phi}/T_{h})^{2}\rangle \equiv F_{z}^{2}$, for transverse and longitudinal fluctuations of the vector $\phi$ field in the lab frame of the collision. By fitting the STAR's experiment data, we extract them as functions of the collision energy, as shown in Fig. \ref{fig:rho-00-eng} (b). These two parameters are almost of the same order, indicating that the anisotropy in the $\phi$ meson's rest frame is mainly induced by its motion relative to the background \cite{Sheng:2023urn}. At lower energies, the fluctuations should be stronger in order to explain the experimental data. The energy dependence can be approximated by analytical formulas $\text{ln} (F_{T,z}^2/m_\pi^2)=a_{T,z}-b_{T,z}\text{ln}(\sqrt{s_\text{NN}}/\text{GeV})$, with the solid and dashed lines in Fig. \ref{fig:rho-00-eng} (a) representing the spin alignment calculated with this analytical expression.

With the extracted parameters, we further predict the azimuthal angle dependence and the rapidity dependence of the $\phi$ meson's spin alignment \cite{Sheng:2022wsy,Sheng:2023urn} . The global spin alignment is nearly a cosin function of the azimuthal angle, and is a growing function of rapidity. Such behaviors are found to be dominated by the broken symmetry due to the motion of the $\phi$ meson relative to the background \cite{Sheng:2023urn}, and are waiting to be checked in experiments in the future.

\section{Spin alignment induced by thermal-shear tensor}\label{sec:shear}

In order to study the shear-induced spin alignment, we start from the Kubo formula, which gives the linear response in a close-to-equilibrium system. For an arbitrary operator $\hat{O}$, the off-equilbrium correction reads
\begin{equation}
    \delta O(x)\equiv \left\langle \hat{O}(x)\right\rangle- 
 \left\langle \hat{O}(x)\right\rangle_{\mathrm{LE}} 
= \partial_{\mu}\beta_{\nu}(x) 
\lim_{K^{\mu}\to 0} 
\frac{\partial}{\partial K_{0}} 
\operatorname{Im} \left[ 
i T(x) 
\int_{-\infty}^{t} d^{4}{x'} 
\left\langle \left[ \hat{O}(x), \hat{T}^{\mu\nu}(x') \right] \right\rangle_{\mathrm{LE}} 
e^{-i K \cdot (x'-x)} 
\right],
\end{equation}
where $\left\langle \hat{O}(x)\right\rangle_{\mathrm{LE}}=\mathrm{Tr
}\left[\rho_{\mathrm{LE}}\hat{O}(x))\right]$ denotes the mean value at local equilibrium, with $\rho_{\mathrm{LE}}$ being the local equilibrium density operator, $\hat{T}^{\mu\nu}$ represents the energy-momentum operator, and $\partial_{\mu}\beta_{\nu}$ is the gradients of the thermal velocity $\beta_{\mu}\equiv
u_\nu/T$. The tensor $\partial_{\mu}\beta_{\nu}$ can be further decomposed into a thermal-shear tensor $\xi_{\mu\nu}=(\partial_{\mu}\beta_{\nu}+\partial_{\nu}\beta_{\mu})/2$ and a thermal vorticity tensor $\varpi_{\mu\nu}=-(\partial_{\mu}\beta_{\nu}-\partial_{\nu}\beta_{\mu})/2$. 
Substituting $\hat{O}$ by the Wigner operator for the vector meson and explicitly evaluating the Kubo formula, we obtain 
\begin{equation}
\int_0^{+\infty} dp_0\,\delta G^{\mu\nu}_<(x,p) \approx 2 T \xi_{\gamma\lambda} \int_0^\infty dp_1^0 \frac{\partial n_B(p_1^0)}{\partial p_1^0} \sum_{a,b=L,T} \rho_a(p_1^0,\mathbf{p}) \rho_b(p_1^0,\mathbf{p}) I^{\mu\nu\gamma\lambda}_{ab}(p_1^0,\mathbf{p},p_1^0,\mathbf{p}),
\end{equation} 
where $\rho_{L,T}(p^0,\mathbf{p})$ denotes the spectral function for the longitudinally (transversely) polarized $\phi$ meson. The tensor $I^{\mu\nu\gamma\lambda}_{ab}$ can be found in Ref. \cite{Dong:2023cng}. Then the spin alignment is calculated by,
\begin{equation}
    \delta \rho_{00}^\phi(x,p)\equiv\rho^\phi_{00}(x,p)-\frac{1}{3} = \frac{L^{\mu\nu}(p_{\rm on}) \int_0^{+\infty} dp_0 \left[ G^<_{\mu\nu}(x,p) + \delta G^<_{\mu\nu}(x,p)\right]}{- \Delta^{\mu\nu}(p_{\rm on}) \int_0^{+\infty} dp_0 \left[ G^<_{\mu\nu}(x,p) + \delta G^<_{\mu\nu}(x,p)\right]}, \label{spin-alignment-shear}
\end{equation}
where the projection operators $L^{\mu\nu}\equiv\epsilon^{\mu\ast}(p_\text{on})\epsilon^\nu(p_\text{on})+\Delta^{\mu\nu}(p_\text{on})/3$ and $\Delta^{\mu\nu}\equiv g^{\mu\nu}-p_\text{on}^\mu p_\text{on}^\nu/m_\phi^2$, with $p_\text{on}^\mu\equiv(E_{\bf p}^\phi,{\bf p})$ being the on-shell momentum.

Under the quasi-particle approximation, the spectral functions are parameterized by a finite width $\Gamma$ and an energy shift $\Delta E$ compared to the energy of a free-streaming particle with the same 3-momentum. In general, $\Gamma$ and $\Delta E$ for the longitudinally polarized meson can differ from those for the transversely polarized meson. Thus we use the subscript $L,T$ to identify these two cases. Then the deviation 
$\delta\rho_{00}$ in Eq. (\ref{spin-alignment-shear}) is simplified to \cite{Dong:2023cng},
\begin{align}
\delta \rho_{00}^\phi(\mathbf{p}) &\approx -\frac{1}{3} \left[ 1 + n_B(E_p^V) \right] \big\{-L_{\mu \nu}(p_{\rm on})\Delta_T^{\mu \nu}(p_{\rm on})C_0(\mathbf{p}) \quad \nonumber \\ 
& + \xi_{\gamma \lambda} L_{\mu \nu}(p_{\rm on}) \Delta_T^{\mu \nu}(p_{\rm on}) \left[ \frac{p_{\rm on}^\gamma p_{\rm on}^\lambda}{(E_p^V)^2} C_1(\mathbf{p}) + \frac{g^{\lambda 0} p_{\rm on}^\gamma + g^{\gamma 0} p_{\rm on}^\lambda - E_p^V g^{\gamma \lambda}}{2E_p^V} (C_T(\mathbf{p}) - C_L(\mathbf{p})) \right] \nonumber \\ 
& + \xi_{\gamma \lambda} L_{\mu \nu}(p_{\rm on}) \left[ \Delta_T^{\gamma \nu}(p_{\rm on}) \Delta_L^{\lambda \mu}(p_{\rm on}) + \Delta_L^{\gamma \nu}(p_{\rm on}) \Delta_T^{\lambda \mu}(p_{\rm on}) \right] C_2(\mathbf{p}) \nonumber\\ 
& + \xi_{\gamma \lambda} L_{\mu \nu}(p_{\rm on}) \left[ \Delta_L^{\gamma \nu}(p_{\rm on}) \Delta_L^{\lambda \mu}(p_{\rm on}) C_L(\mathbf{p}) + \Delta_T^{\gamma \nu}(p_{\rm on}) \Delta_T^{\lambda \mu}(p_{\rm on}) C_T(\mathbf{p}) \right] \Bigg\}\,,\label{delta-rho-final}
\end{align}
which can be expressed as
\begin{equation}
\delta \rho_{00}^\phi(\mathbf{p})=\delta \rho_{00}^{(\xi=0)}(\mathbf{p}) +\xi_{\mu\nu}C^{\mu\nu}(\bf p)\,. \label{Cmunu}
\end{equation}
Here the first term on the right-hand-side is induced by the
motion of meson relative to the thermal background, and the remaining part is the linear response to the thermal shear-tensor. It is noted that there is no linear response to thermal vorticity tensor. In Eq. (\ref{delta-rho-final}), $C_0$, $C_1$, $C_2$, and $C_{T/L}$ are dimensionless coefficients related to widths and energy shifts of spectral functions, which are given by
\begin{align}
C_0 &= \frac{1 + n_B (E_p^V) + \frac{T}{E_p^V} (\Delta E_T - \Delta E_L)}{1 + n_B (E_p^V)}, \\
C_1 &= \frac{(E_p^V)^2}{m_V} \left( \frac{1}{\Gamma_T} - \frac{1}{\Gamma_L} \right) + n_B (E_p^V) \frac{(E_p^V)^2}{m_V T} \left( \frac{\Delta E_L}{\Gamma_L} - \frac{\Delta E_T}{\Gamma_T} \right), \\
C_2 &= \frac{4 m_V E_p^V (\Gamma_L \Delta E_T + \Gamma_T \Delta E_L)}{4 (E_p^V)^2 (\Delta E_T - \Delta E_L)^2 + m_V^2 (\Gamma_L + \Gamma_T)^2}, \\
C_{T/L} &= \frac{2 E_p^V \Delta E_{T/L}}{m_V \Gamma_{T/L}}\,.
\end{align}

\begin{figure}[h]
    \centering
    \includegraphics[width=0.42\linewidth]{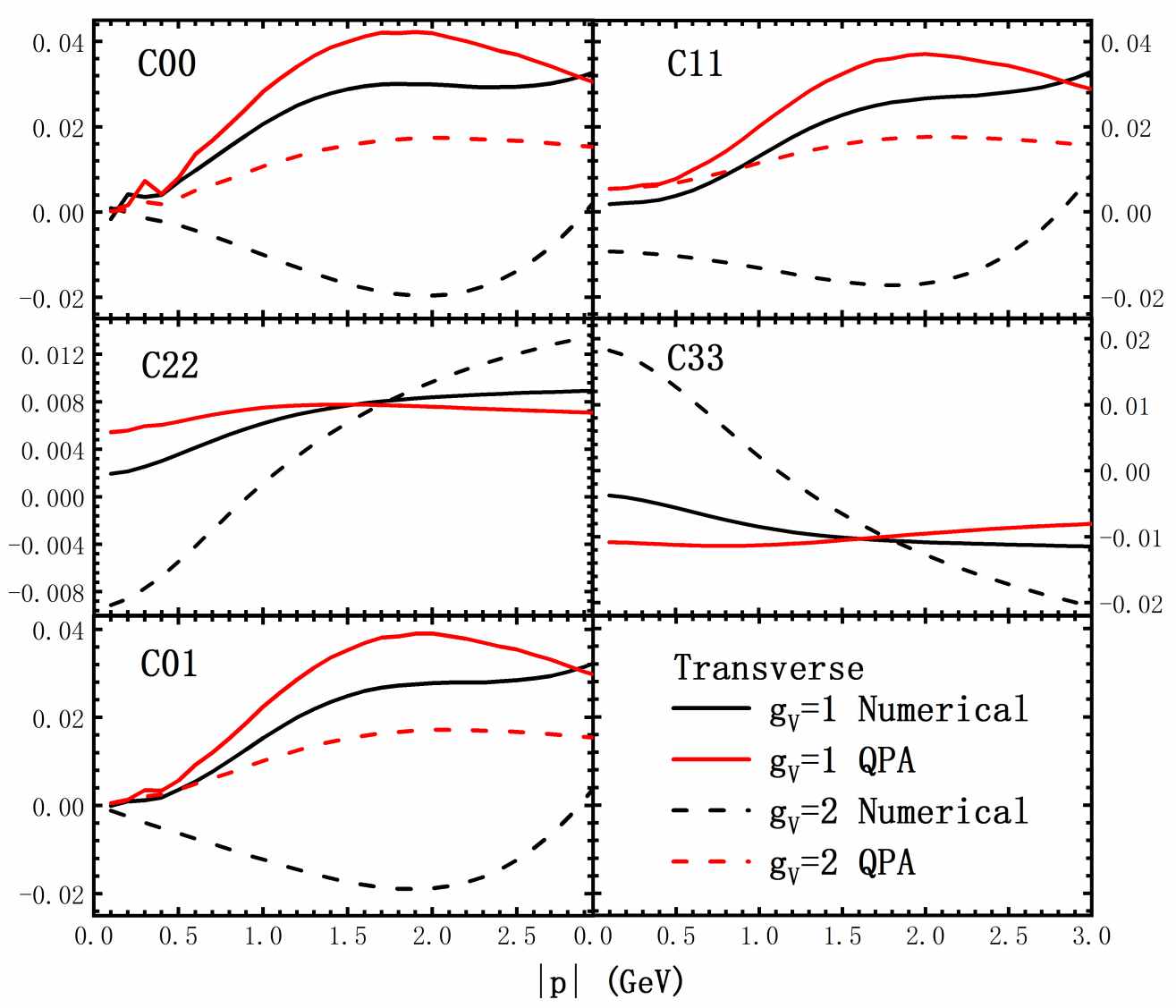}
    \includegraphics[width=0.43\linewidth]{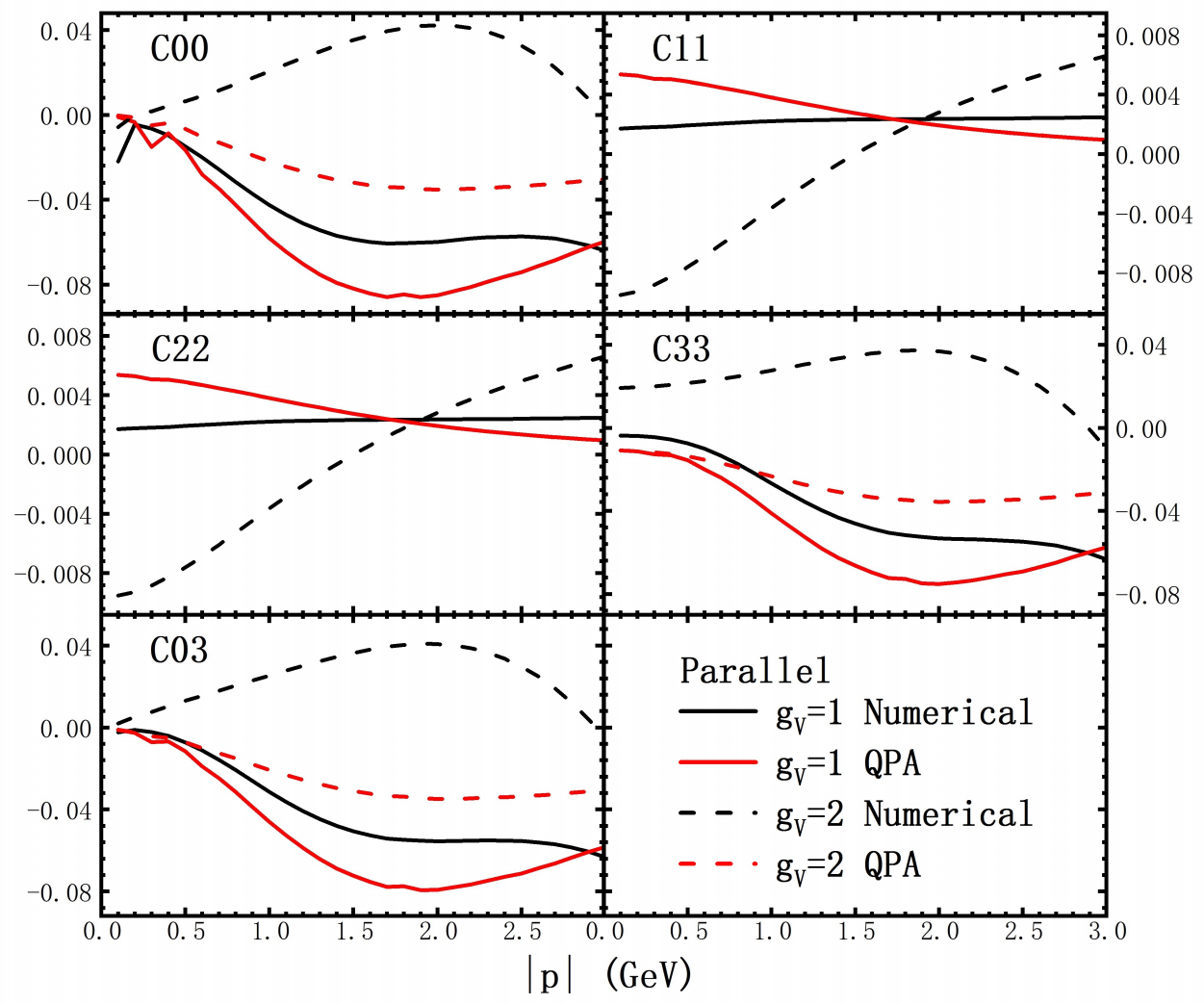}
    \caption{The numerical results for $C^{\mu\nu}$ in Eq. (6) where the directions of the spin quantization are parallel and perpendicular to the momentum \cite{Dong:2023cng}. }
    \label{fig:enter-label}
\end{figure}

Considering a quark-meson interaction, we numerically evaluated the width and energy shifts for the $\phi$ meson. Then different components of the coefficient $C^{\mu\nu}$ for the shear-induced spin alignment, appearing in Eq. (\ref{Cmunu}), are shown in Fig. \ref{fig:enter-label} as functions of the momentum. Here we choose the temperature $T=158.4$ MeV and the chemical potential for $s$ quark $\mu_s=7.4$ MeV, corresponding to the freezeout conditions in heavy-ion collisions at $\sqrt{s_\text{NN}}=200$ GeV. We take two special cases for the spin quantization direction, one is parallel to the momentum and the other one is perpendicular. We show in Fig. \ref{fig:enter-label} the results calculated with or without the quasi-particle approximation, labeled by {\it QPA} and {\it Numerical}, respectively. For the whole considered momentum region, components of $C^{\mu\nu}$ is of the order of  $10^{-2}$ or smaller. Therefore if the thermal shear tensor in heavy-ion collisions is $10^{-2}$, its contribution to the spin alignment will be $10^{-4}$ or even smaller \cite{Dong:2023cng}. 

\section{Summary}

In this proceeding we briefly review our understandings on the spin alignment of the $\phi$ meson, especially on contributions of the strong field and the thermal shear tensor. Main results from relativistic quark coalescence model with spin degrees of freedom are listed, presenting the relation between meson's spin alignment and spin polarizations of its constituent quark and antiquark. We then discussed the shear-induced spin alignment, which is calculated based on the Kubo formula in a quark-meson model. The magnitude of the shear-induced part is found to be very small, while the strong field can successfully explain the spin alignment observed by the STAR collaboration. The validity of our model with strong field fluctuation needs to be checked in future experiments through the azimuthal angle or rapidity dependence of the $\phi$ meson's spin alignment.

\bibliographystyle{elsarticle-num}
\bibliography{cas-refs}

\begin{thebibliography}{10}
\expandafter\ifx\csname url\endcsname\relax
  \def\url#1{\texttt{#1}}\fi
\expandafter\ifx\csname urlprefix\endcsname\relax\def\urlprefix{URL }\fi
\expandafter\ifx\csname href\endcsname\relax
  \def\href#1#2{#2} \def\path#1{#1}\fi

\bibitem{Liang:2004ph}
Z.-T. Liang, X.-N. Wang, {Globally polarized quark-gluon plasma in non-central A+A collisions}, Phys. Rev. Lett. 94 (2005) 102301, [Erratum: Phys.Rev.Lett. 96, 039901 (2006)].
\newblock \href {http://arxiv.org/abs/nucl-th/0410079} {\path{arXiv:nucl-th/0410079}}, \href {https://doi.org/10.1103/PhysRevLett.94.102301} {\path{doi:10.1103/PhysRevLett.94.102301}}.

\bibitem{STAR:2017ckg}
L.~Adamczyk, et~al., {Global $\Lambda$ hyperon polarization in nuclear collisions: evidence for the most vortical fluid}, Nature 548 (2017) 62--65.
\newblock \href {http://arxiv.org/abs/1701.06657} {\path{arXiv:1701.06657}}, \href {https://doi.org/10.1038/nature23004} {\path{doi:10.1038/nature23004}}.

\bibitem{Liang:2004xn}
Z.-T. Liang, X.-N. Wang, {Spin alignment of vector mesons in non-central A+A collisions}, Phys. Lett. B 629 (2005) 20--26.
\newblock \href {http://arxiv.org/abs/nucl-th/0411101} {\path{arXiv:nucl-th/0411101}}, \href {https://doi.org/10.1016/j.physletb.2005.09.060} {\path{doi:10.1016/j.physletb.2005.09.060}}.

\bibitem{STAR:2022fan}
M.~S. Abdallah, et~al., {Pattern of global spin alignment of \ensuremath{\phi} and K$^{*0}$ mesons in heavy-ion collisions}, Nature 614~(7947) (2023) 244--248.
\newblock \href {http://arxiv.org/abs/2204.02302} {\path{arXiv:2204.02302}}, \href {https://doi.org/10.1038/s41586-022-05557-5} {\path{doi:10.1038/s41586-022-05557-5}}.

\bibitem{Yang:2017sdk}
Y.-G. Yang, R.-H. Fang, Q.~Wang, X.-N. Wang, {Quark coalescence model for polarized vector mesons and baryons}, Phys. Rev. C 97~(3) (2018) 034917.
\newblock \href {http://arxiv.org/abs/1711.06008} {\path{arXiv:1711.06008}}, \href {https://doi.org/10.1103/PhysRevC.97.034917} {\path{doi:10.1103/PhysRevC.97.034917}}.

\bibitem{Sheng:2020ghv}
X.-L. Sheng, Q.~Wang, X.-N. Wang, {Improved quark coalescence model for spin alignment and polarization of hadrons}, Phys. Rev. D 102~(5) (2020) 056013.
\newblock \href {http://arxiv.org/abs/2007.05106} {\path{arXiv:2007.05106}}, \href {https://doi.org/10.1103/PhysRevD.102.056013} {\path{doi:10.1103/PhysRevD.102.056013}}.

\bibitem{Sheng:2022wsy}
X.-L. Sheng, L.~Oliva, Z.-T. Liang, Q.~Wang, X.-N. Wang, {Spin Alignment of Vector Mesons in Heavy-Ion Collisions}, Phys. Rev. Lett. 131~(4) (2023) 042304.
\newblock \href {http://arxiv.org/abs/2205.15689} {\path{arXiv:2205.15689}}, \href {https://doi.org/10.1103/PhysRevLett.131.042304} {\path{doi:10.1103/PhysRevLett.131.042304}}.

\bibitem{Sheng:2022ffb}
X.-L. Sheng, L.~Oliva, Z.-T. Liang, Q.~Wang, X.-N. Wang, {Relativistic spin dynamics for vector mesons}, Phys. Rev. D 109~(3) (2024) 036004.
\newblock \href {http://arxiv.org/abs/2206.05868} {\path{arXiv:2206.05868}}, \href {https://doi.org/10.1103/PhysRevD.109.036004} {\path{doi:10.1103/PhysRevD.109.036004}}.

\bibitem{Li:2022vmb}
F.~Li, S.~Y.~F. Liu, {Tensor polarization and the dissipative damping of vector meson in QCD Medium} (6 2022).
\newblock \href {http://arxiv.org/abs/2206.11890} {\path{arXiv:2206.11890}}.

\bibitem{Wagner:2022gza}
D.~Wagner, N.~Weickgenannt, E.~Speranza, {Generating tensor polarization from shear stress}, Phys. Rev. Res. 5~(1) (2023) 013187.
\newblock \href {http://arxiv.org/abs/2207.01111} {\path{arXiv:2207.01111}}, \href {https://doi.org/10.1103/PhysRevResearch.5.013187} {\path{doi:10.1103/PhysRevResearch.5.013187}}.

\bibitem{Dong:2023cng}
W.-B. Dong, Y.-L. Yin, X.-L. Sheng, S.-Z. Yang, Q.~Wang, {Linear response theory for spin alignment of vector mesons in thermal media}, Phys. Rev. D 109~(5) (2024) 056025.
\newblock \href {http://arxiv.org/abs/2311.18400} {\path{arXiv:2311.18400}}, \href {https://doi.org/10.1103/PhysRevD.109.056025} {\path{doi:10.1103/PhysRevD.109.056025}}.

\bibitem{Becattini:2022zvf}
F.~Becattini, {Spin and polarization: a new direction in relativistic heavy ion physics}, Rept. Prog. Phys. 85~(12) (2022) 122301.
\newblock \href {http://arxiv.org/abs/2204.01144} {\path{arXiv:2204.01144}}, \href {https://doi.org/10.1088/1361-6633/ac97a9} {\path{doi:10.1088/1361-6633/ac97a9}}.

\bibitem{Chen:2023hnb}
J.~Chen, Z.-T. Liang, Y.-G. Ma, Q.~Wang, {Global spin alignment of vector mesons and strong force fields in heavy-ion collisions}, Sci. Bull. 68 (2023) 874--877.
\newblock \href {http://arxiv.org/abs/2305.09114} {\path{arXiv:2305.09114}}, \href {https://doi.org/10.1016/j.scib.2023.04.001} {\path{doi:10.1016/j.scib.2023.04.001}}.

\bibitem{Becattini:2024uha}
F.~Becattini, M.~Buzzegoli, T.~Niida, S.~Pu, A.-H. Tang, Q.~Wang, {Spin polarization in relativistic heavy-ion collisions}, Int. J. Mod. Phys. E 33~(06) (2024) 2430006.
\newblock \href {http://arxiv.org/abs/2402.04540} {\path{arXiv:2402.04540}}, \href {https://doi.org/10.1142/S0218301324300066} {\path{doi:10.1142/S0218301324300066}}.

\bibitem{Chen:2024afy}
J.-H. Chen, Z.-T. Liang, Y.-G. Ma, X.-L. Sheng, Q.~Wang, {Vector meson\textquoteright{}s spin alignments in high energy reactions}, Sci. China Phys. Mech. Astron. 68~(1) (2025) 211001.
\newblock \href {http://arxiv.org/abs/2407.06480} {\path{arXiv:2407.06480}}, \href {https://doi.org/10.1007/s11433-024-2495-1} {\path{doi:10.1007/s11433-024-2495-1}}.

\bibitem{Sheng:2019kmk}
X.-L. Sheng, L.~Oliva, Q.~Wang, {What can we learn from the global spin alignment of $\phi$ mesons in heavy-ion collisions?}, Phys. Rev. D 101~(9) (2020) 096005, [Erratum: Phys.Rev.D 105, 099903 (2022)].
\newblock \href {http://arxiv.org/abs/1910.13684} {\path{arXiv:1910.13684}}, \href {https://doi.org/10.1103/PhysRevD.101.096005} {\path{doi:10.1103/PhysRevD.101.096005}}.

\bibitem{Sheng:2023urn}
X.-L. Sheng, S.~Pu, Q.~Wang, {Momentum dependence of the spin alignment of the \ensuremath{\phi} meson}, Phys. Rev. C 108~(5) (2023) 054902.
\newblock \href {http://arxiv.org/abs/2308.14038} {\path{arXiv:2308.14038}}, \href {https://doi.org/10.1103/PhysRevC.108.054902} {\path{doi:10.1103/PhysRevC.108.054902}}.

\end{thebibliography}

\end{document}